# Towards Trustworthy Artificial Intelligence for Equitable Global Health

Hong Qin[1,*], Jude Kong[2,*], Wandi Ding[3], Ramneek Ahluwalia[4], Christo El Morr[2], Zeynep Engin[5], Jake Okechukwu Effoduh[2], Rebecca Hwa[6], Serena Jingchuan Guo[7], Laleh Seyyed-Kalantari[2], Sylvia Kiwuwa Muyingo[8], Candace Makeda Moore[9], Ravi Parikh[10], Reva Schwartz[11], Dongxiao Zhu[12], Xiaoqian Wang[13], Yiye Zhang[14]

1 University of Tennessee at Chattanooga, U.S.A.
2 York University, Canada
3 Middle Tennessee State University, U.S.A.,
4 Higher Health, South Africa
5 Data for Policy CIC, UK.
6 University of Pittsburgh, U.S.A.
7 University of Florida, U.S.A.
8 African Population and Health Research Center, Kenya
9 Netherlands eScience Center, Netherlands
10 University of Pennsylvania, U.S.A.
11 National Institute of Standards and Technology, U.S.A.
12 Wayne State University, U.S.A.
13 Purdue University, U.S.A.
14 Weill Cornell Medicine, U.S.A.
*Correspondence should be addressed to hong-qin@utc.edu and dkong@yorku.ca.

**Abstract**

Artificial intelligence (AI) can potentially transform global health, but algorithmic bias can exacerbate social inequities and disparity. Trustworthy AI entails the intentional design to ensure equity and mitigate potential biases. To advance trustworthy AI in global health, we convened a workshop on Fairness in Machine Intelligence for Global Health (FairMI4GH). The event brought together a global mix of experts from various disciplines, community health practitioners, policymakers, and more. Topics covered included managing AI bias in socio-technical systems, AI's potential impacts on global health, and balancing data privacy with transparency. Panel discussions examined the cultural, political, and ethical dimensions of AI in global health. FairMI4GH aimed to stimulate dialogue, facilitate knowledge transfer, and spark innovative solutions. Drawing from NIST's AI Risk Management Framework, it provided suggestions for handling AI risks and biases. The need to mitigate data biases from the research design stage, adopt a human-centered approach, and advocate for AI transparency was recognized. Challenges such as updating legal frameworks, managing cross-border data sharing, and motivating developers to reduce bias were acknowledged. The event emphasized the necessity of diverse viewpoints and multi-dimensional dialogue for creating a fair and ethical AI framework for equitable global health.

**Fairness and Trust in AI in Post-COVID-19 Era**

The COVID-19 pandemic highlighted the critical need for Artificial Intelligence (AI) within the realm of global health. As the global community grappled with the health crisis, AI emerged as an invaluable tool, enabling disease tracking, predictive modeling of virus spread, resource optimization, assistance in medical diagnoses, and even contributing to expedited vaccine development (Syrowatka et al 2021). However, as AI took center stage in the battle against the pandemic, it concurrently unveiled the intricate complexities associated with fairness within these systems.

Fairness in AI is not a mere technical concern but a deep-seated issue of ethics and societal equity. The effectiveness of AI-driven interventions, particularly amidst crises, is not solely dependent on algorithmic sophistication or advanced data modeling, but fundamentally hinges on the trust that these systems engender within the public. This trust is fundamentally influenced by the perceived fairness, transparency, and accountability inherent in AI technologies. As such, the pandemic served as a stark revelation of a critical paradigm: social trust in AI is not an optional add-on but a cardinal necessity.

Additionally, the crisis shed light on the disparate impact of AI, with its benefits and drawbacks unevenly distributed across societal strata, potentially exacerbating existing health inequities. Such disparities often originate from underlying biases embedded within AI systems. Consequently, public trust in AI is intrinsically tied to the successful incorporation of fairness principles into AI algorithms and practices, to ensure equitable distribution of benefits and mitigation of harm, particularly for the most vulnerable. This revelation, anchored in fairness and trust, signifies a defining moment in the evolution of AI application and development in the post-pandemic world.

**Some Case Studies from the COVID-19 Pandemic**

The COVID-19 pandemic has highlighted the powerful role of AI in healthcare, including predicting virus spread, optimizing resources, aiding diagnoses, and accelerating vaccine development (Syrowatka et al 2021). However, several cases during the crisis also revealed the significant implications of fairness within AI systems.

The Veterans Health Administration uses the Care Assessment Needs (CAN) score to predict mortality risk and enhance resource allocation among over 5 million Veterans; however, a study found evidence of racial bias in the model, with mortality risks being underestimated for Black compared to White Veterans (Parikh et al, 2021). The identified disparities, likely influenced by confounded social factors, suggest statistical unfairness, and efforts to address this through class imbalance corrections have shown only marginal improvements in fairness.

During the implementation of an AI model for postpartum depression prediction at NewYork-Presbyterian Hospital/Weill Cornell Medicine, the researchers prioritized addressing algorithmic bias. The evaluation of fairness metrics placed special emphasis on race as a protected attribute, ensuring that predictions and interventions were equitable across diverse racial groups (Zhang et al, 2020).

Another case involved an AI model predicting opioid overdose risks using insurance claims data, which showed racial bias (Guo et al unpublished results). This led to a deeper investigation into the root causes of this bias and subsequent efforts to mitigate it through strategies such as data rebalancing, removing sensitive attributes, and applying penalized parameters. This study highlighted the challenge in ensuring not only the fairness of machine learning models but also the overall fairness of the system and outcomes, requiring a careful balance between prediction accuracy and equitable representation.

During the pandemic, machine learning was extensively used for triage and timely diagnosis of symptomatic cases, forecasting disease dynamics, adjusting epidemic curves, and predicting patients' illness trajectories and responses to treatment (Syrowatka et al, 2021). However, social equity and algorithmic fairness often lacked sufficient attention.

One notable example is AI's deployment in telemedicine during the pandemic (Melore, 2021). AI chatbots, used extensively for patient screening and triage, were found to lack diversity in their training data, leading to racial and gender biases (Heath, 2023; Hundt et al 2022). For instance, darker-skinned and female chatbot avatars received significantly more negative user evaluations (Cadogan, 2023).

These experiences underscore the need for diversity in AI training data and the necessity of checks and balances to identify and mitigate algorithmic biases. As we move toward an increasingly AI-driven healthcare future, fairness in AI is not just desirable but an essential imperative.

**Understanding the Bias in AI for Global Health**

In data science, the terms measurement error, inaccuracy, and bias signify distinct but interconnected issues that can emerge within a dataset. Measurement error generally refers to the gap between the true value and the measured value, a discrepancy that can arise from various factors such as human error, equipment inaccuracies, or inherent variability in the measured entity. Inaccuracy is the extent of deviation from the true value, a result of either measurement errors or issues with the data collection process such as misclassification, data omission, or errors in data entry. Bias is the concept that warrants the most attention in the context of AI and global health. AI bias can refer to systematic errors that result in over- or underestimation of a value, due to flawed data collection methods, selection bias, confounding variables, or even how data are processed or analyzed.

During the workshop, Dr. Candace Makeda Moore brings attention to the risk of AI applications amplifying societal hierarchies and biases, particularly within the healthcare sector. Drawing on her personal experiences and insights, she notes that AI-induced harm is frequently non-random and can often be interpreted through paradigms such as the social dominance theory. She emphasizes the necessity for a diverse team in the development of algorithms, a measure that would prevent harmful stereotypes and social class disparities from being encoded into AI systems. In the same vein, she advocates for regular monitoring of deployed algorithms and encourages fair and open practices.

Dr. Xiaoqian Wang stresses the importance of understanding the reasoning behind machine learning predictions. Advocating for transparency and interpretability in machine learning models, she introduces a three-pronged approach involving pre-processing, in-processing, and post-processing to effectively mitigate potential biases (Chai et al, 2022).

Addressing these issues necessitates careful scrutiny of data collection and analysis processes. Data scientists often resort to error analysis, validation with ground-truth data, cross-validation, and bias correction methods to fortify machine learning models. Despite these measures, several outstanding challenges remain in the field, including the limitations and biases inherent in data gathering and sharing, the pressing need for trustworthy AI, and the risk of AI reinforcing existing digital and socioeconomic disparities. To counteract these challenges, potential solutions such as federated learning for data privacy, label smoothing for fairness assurance without relying on demographic data, and generative AI chatbots for secure data collection are suggested.

Current methods to address bias include techniques such as normalization, regularization imposition on models, and label smoothing. Future research areas that show promise include exploring the intersection of individual and group fairness, improving scalability and automation in bias identification, and effectively capturing socioeconomic features at an individual level.

The predominant focus on high-income countries in the collection and analysis of global health data can magnify existing health inequities. The current concentration of research in high-income settings also limits the generalizability of these tools. Hence, we need to expand data collection and research efforts to include low- and middle-income countries, ensuring a more accurate representation of global health scenarios.

The workshop dialogues and discussions underscore the urgent need for a multifaceted approach in constructing trustworthy AI systems that foster health equity. This comprehensive approach encompasses addressing technical and infrastructural issues, formulating effective policies, ensuring data representativeness, and involving diverse stakeholders in the decision-making processes. As highlighted by both Dr. Moore and Wang, collaboration, transparency, and a steadfast commitment to fairness are the cornerstones in this endeavor.

**Unraveling the Intricacies: Privacy, Bias, Transparency and Fairness.**

The workshop was notably marked by two recurrent themes–- the delicate balance between open data and data privacy, and the convoluted problem of understanding and rectifying biases within AI systems. AI systems, particularly in healthcare, derive their efficacy from the richness and diversity of the data employed for training. To ensure fairness and unbiased performance, these models necessitate access to data that is representative of the broader population. This fundamental requirement, however, clashes directly with the indispensable need for privacy protection, a cornerstone of ethical AI deployment.

To navigate this complex landscape, the idea of federated learning was discussed. Federated learning permits model training across decentralized devices or servers holding local data samples, negating the need to share raw data. Despite its potential, the universal application of federated learning is far from straightforward. It demands substantial infrastructural capabilities and broad acceptance across various stakeholders. Moreover, federated learning does not dissolve concerns related to bias, as localized data could misrepresent or underrepresent certain population segments.

The inherent opacity of AI systems, which conceals their internal operations from users, complicates understanding how biases originate within these systems and how they can be rectified. The issue is accentuated when biased AI decisions lead to tangible real-world repercussions, such as skewed resource allocation during a pandemic.

The workshop participants recognized a vital need for publicly accessible, user-friendly tools, allowing non-technical users to understand and scrutinize AI systems. These AI tools should be equipped to tackle a variety of bias types, spanning from race, gender, to intersectionality. The significance of bias prediction models and sensitivity checks to create more representative and unbiased AI systems was also emphasized.

The workshop participant understood that the creation of such tools and the incorporation of these practices are not trivial tasks. These tools involve not just technical prowess but also a comprehension of the social, cultural, and ethical implications of AI systems, which demands the participation of diverse stakeholders, including AI researchers, policymakers, ethicists, and the public.

In essence, the significant challenges of reconciling privacy with open data and understanding bias within AI systems, though substantial, are not insurmountable. Conversations during the panel discussions indicate that through collaborative efforts, innovative approaches, and an unwavering commitment to fairness, we can navigate these complexities towards a future of fair, trustworthy AI in global health.

**Conclusions and the Road Ahead**

Our journey towards fair AI, particularly in healthcare, as outlined by the National Institute of Standards and Technology (NIST), is both complex and necessary. The AI Risk Management Framework (AI RMF) by NIST offers a structured approach to manage risks and biases in AI systems, prioritizing real-world impacts and fostering a culture of responsible AI practices. (NIST 2023) To mitigate biases – computational, systemic, and human cognitive – and other AI risks, the framework encourages a socio-technical perspective, emphasizing activities that center the human instead of the AI system, testing, evaluation, validation and verification (TEVV) practices across the AI lifecycle, AI governance, and consideration of societal impacts.

Recognizing AI as a potent tool for global health, we need to grapple with technical challenges, craft effective policies, assure data representativeness, and engage diverse stakeholders in understanding the risks of AI to enhance organizational decision-making. The COVID-19 pandemic has reinforced that this undertaking is not merely necessary but imperative for global societal welfare.

Key lessons learned include the importance of identifying and mitigating data biases from the research design stage onwards, the necessity of a human-first approach in AI development, and the critical role of ethical considerations and human rights and societal values in this process. Bias mitigation, while challenging, is crucial, and so is the need for AI transparency and interoperability to prevent potential harms and amplification of historical biases.

Best practices suggest the need for public accessibility of tools and frameworks for bias identification and mitigation, promotion of bias prediction models, the incorporation of legal mechanisms for AI fairness such as debiasing orders and algorithmic impact assessments, and fostering trust through AI transparency.

Many challenges persist. Existing legal frameworks often lag behind the rapid evolution of AI technologies. Global data sharing and cooperation face obstacles due to varying national laws and standards. Incentivizing developers to reduce bias in their systems requires innovative approaches. Addressing these challenges involves employing AI transparency tools, implementing bias-check frameworks, and leveraging open-source AI models to democratize access.

The FairMI4GH workshop convened a globally representative, multidisciplinary group of participants, serving as a rich platform for varied perspectives on AI development. This mix of technical, legal, and ethical viewpoints helped uncover various aspects of AI in global health, emphasizing the significance of diverse voices, as supported by others (Frehywot et al 2023). Such diversity is critical to crafting AI systems that are fair, ethical, transparent, and beneficial to all. Participants, including experts from different disciplines, community health practitioners, program managers, governmental agency staff scientists, and other stakeholders from Africa, Europe, North America, Asia, Latin America, the Caribbean, and the Middle East and North Africa region, brought their unique perspectives to the table. Notably, regional differences in approach and focus emerged, with African representatives prioritizing capacity-building and infrastructure development, while those from the USA focusing on the balance between privacy and transparency in AI applications. These discussions highlighted a global trend towards more context-specific, ethically-conscious, and inclusive AI solutions in global health.

The road ahead is challenging but promising. By employing comprehensive frameworks like the NIST's AI RMF, fostering global collaboration, and committing to equitable AI, we can navigate towards a future of trustworthy, responsible, and inclusive AI.

**Acknowledgement**
Much of these contents were based on a workshop organized in April 2023. We thank the staff of Consortium for University or Global Health (CUGH) for logistical support in organizing the workshop. We acknowledge the assistance of AI tools in processing recorded discussions. HQ thanks the support of USA National Science Foundation (NSF) award 2200138. JDK acknowledges support from Canada's International Development Research Centre (IDRC) (Grant No. 109981).